\documentstyle[12pt]{article}
\font\double=msym10

\def\cc{{\hbox{\double C}}}
\def\nn{{\hbox{\double N}}}

\def\aa{{\cal A}}
\def\dd{{\cal D}}

\def\hh{{\cal H}}

\def\t{{\rm tr}}

\def\lb{\left[}
\def\rb{\right]}
\def\lp{\left(}
\def\rp{\right)}
\def\la{\left\{}
\def\ra{\right\}}

\def\ot{\otimes}
\def\op{\oplus}

\def\bb{\begin{eqnarray}}
\def\ee{\end{eqnarray}}
\def\eee{\nonumber\end{eqnarray}}
\def\pp{\pmatrix}

\begin{document}

\hsize 17truecm
\vsize 24truecm
\font\twelve=cmbx10 at 13pt
\font\eightrm=cmr8
\baselineskip 18pt

\begin{titlepage}

\centerline{\twelve CENTRE DE PHYSIQUE THEORIQUE}
\centerline{\twelve CNRS - Luminy, Case 907}
\centerline{\twelve 13288 Marseille Cedex}
\vskip 4truecm

\centerline{\twelve A LEFT-RIGHT SYMMETRIC MODEL}
\centerline{\twelve  A LA CONNES-LOTT}

\bigskip

\begin{center}
{\bf Bruno IOCHUM}
\footnote{ and Universit\'e de Provence} \\
\bf Thomas SCH\"UCKER $^{1}$
\end{center}

\vskip 2truecm
\leftskip=1cm
\rightskip=1cm
\centerline{\bf Abstract}

\medskip

We present a left-right symmetric model with gauge
group $U(2)_L\times U(2)_R$ in the Connes-\-Lott
non-\-commutative frame work. Its gauge symmetry is
broken spontaneously, parity remains unbroken.

\vskip 2truecm

\vskip 3truecm

\noindent December 1993

\noindent CPT-93/P.2973

 \end{titlepage}

A Yang-\-Mills-\-Higgs model is specified by the
choice of a Lie group $G$, representations $\phi,\
\psi_L,\ \psi_R$ for scalars, left and right handed
fermions, and a Higgs potential $V(\phi)$.
 From these data, one then computes the mass matrices
(i.e. masses and mixing angles) of the gauge bosons,
fermions and scalars. In the Connes-Lott scheme [1], the
input consists of an involution algebra $\aa$ (which
contains the group $G$), of (algebra) representations
$\psi_L,\ \psi_R$ and of the fermion mass matrix. The
rest, that is the boson mass matrix and the complete
Higgs sector, is output. So far only few models have
been computed in detail with the Connes-Lott
algorithm, the standard model $G=SU(3) \times SU(2)
\times U(1)$ [1,2,3], a ``chiral version'' of
electromagnetism $G=U(1)$ [1], a left-\-right
symmetric model with $G=U(1)_L \times U(1)_R$ [1] and
a model with $G=U(2)\times U(1)$ [4]. The purpose of
the present work is twofold. To get acquainted with the
fine points of the Connes-Lott algorithm it seems
indispensable to study further examples. Secondly, the
most striking success of the Connes-Lott scheme is
certainly the geometric explanation of the spontaneous
break down of gauge symmetry. It natural to ask
whether parity violation can be explained at the same
time. To this end, we study the left-\-right symmetric
$U(2)_L\times U(2)_R$ model. As a warm up, we also
consider a $U(2)$ model. To get started let us briefly
summarize our notations [4].

\section{Notations}

Let $\aa$ be a finite dimensional involution algebra
(associative, with unit 1 and involution $^*$) and
$\rho$ a faithful representation of $\aa$ on a finite
dimensional Hilbert space $\hh$.
Let $\chi$, ``the chirality'', be a self adjoint operator
on $\hh$, with $\chi^2 =1$ and let $\dd$, ``the
(internal) Dirac operator'', be another self adjoint
operator on $\hh$. Furthermore
we suppose that $\rho (a) $ is even:
\bb    \rho(a) \chi =
\chi \rho(a) \eee
 for all $a \in \aa$ and that $\dd $ is
odd:
\bb    \dd \chi = -\chi \dd .\eee
In other
words the representation $\rho$ is reducible and
decomposes into a left handed and a right handed part
$\rho_L$ and $\rho_R$  living on the left handed and
right handed Hilbert spaces
\bb   \hh_L :=
{{1-\chi}\over2} \hh,\eee
\bb   \hh_R := {{1+\chi}\over2}
\hh.\eee
We can always
pick a basis such that
\bb    \chi = \pmatrix {1_L&0\cr
0&-1_R}.\eee
Then
\bb   \rho = \pmatrix {\rho_L & 0 \cr 0&
\rho_R},\eee
 \bb   {\dd}= \pmatrix {0 & M \cr
 M^\ast & 0},\eee
with $M$ a matrix of size
$\rm{dim}\hh_L \times \rm{dim}\hh_R$, ``the mass
matrix''. The triple $(\hh,\chi,\dd)$ plays an
important role in non-\-commutative geometry where
it is called K-cycle.

The representation $\rho$ is extended from the
algebra $\aa$ to its universal differential envelop
 $\hat\Omega\aa$ by
$$\pi(a_0\delta a_1...\delta a_p) :=
(-i)^p\rho(a_0)[\dd,\rho(a_1)] ...[\dd,\rho(a_p)].$$
Although $\rho$ is faithful, $\pi$ is not. The central
piece of Connes' theory is the differential algebra
\bb   \Omega\aa =\bigoplus_{p\in\nn}
\Omega^p\aa\nonumber\ee
defined by
\bb \Omega^0\aa:=\rho(\aa),\nonumber\ee
\bb \Omega^1\aa:=\pi(\hat\Omega^1\aa),\nonumber\ee
\bb   \Omega^p\aa :=
{{\pi(\hat\Omega^p\aa)}\over
{\pi(\delta(\ker\pi)^{p-1})}},\quad p\geq2.\nonumber\ee
The involution $^*$ is extended to the universal
differential envelop $\hat \Omega\aa$ by
\bb   (\delta a)^* := \delta(a^*) =:\delta a^*,
\quad a\in \aa\eee
and passes to the quotient $\Omega\aa$. We warn the
reader that the same symbols, $\delta$ for the
differential and $^*$ for the involution, are used in
both differential algebras $\hat\Omega\aa$ and
$\Omega\aa$.

Since the elements of
$\pi(\hat\Omega\aa)$ are operators on the Hilbert
space $\hh$, i.e. concrete matrices, they have a
natural scalar product defined by
\bb   <\hat\phi,\hat\psi> := \t (\hat\phi^*\hat\psi),
\quad  \hat\phi, \hat\psi \in
\pi(\hat\Omega^p\aa)\eee
   for forms of equal degree
and by zero for the scalar product of two  forms of
different degree. With this scalar product, the quotient
$\Omega\aa$ is a subspace of $\pi(\hat\Omega\aa)$,
the one orthogonal to the ``junk'' $J :=
\delta\ker\pi$. As a subspace $\Omega\aa$  inherits a
scalar product which deserves a special name ( , ). It is
given by
\bb   (\phi,\psi) =
\t(\phi^*P\psi), \quad \phi, \psi \in \Omega^p\aa\eee
where $P$ is the orthogonal projector in
$\pi(\hat\Omega\aa)$  onto the ortho-\-complement of
$J$ and $\phi$ and $\psi$ are any  representatives in
their classes. Again the scalar product vanishes  for
forms with different degree.

A Higgs (multiplet) or gauge potential $H$ is by
definition an antihermitian element of
$\Omega^1\aa$. The Higgses carry an affine
representation of the group of unitaries
\bb    G = \{g\in \aa,\ gg^\ast
=g^*g=1\} \eee
defined by
\bb H^g &:=&\
\rho(g)H\rho(g^{-1})+\rho(g)\delta \rho(g^{-1}) \cr
       &=&\ \rho(g)H\rho(g^{-1})+(-i)\rho(g)[\dd,\rho
(g^{-1})] \cr
&=&\
\rho(g)(H-i\dd)\rho(g^{-1})+i\dd.\label{gauge}\ee
 $H^g$ is
the ``gauge transformed of $H$''.  To motivate the term
gauge potential, we note that every $H$ defines a
covariant derivative  $\delta +H$. This covariant
derivative operates on $\Omega\aa$:
\bb   ^g\psi :=
\rho(g)\psi, \quad \psi\in\Omega\aa\eee
 and is covariant
under the left action of $G$ :
\bb   (\delta+H^g)\
^g\psi = \ ^g\lb(\delta+H)\psi\rb.\eee
 As usual we define the
curvature $C$ of $H$ by
\bb   C := \delta H+H^2\ \in
\Omega^2\aa.\eee
Note that here and later $H^2$ is considered as element
of $\Omega^2\aa$ which means it is the projection $P$
applied to $H^2\in \pi(\hat\Omega^2\aa)$.
The curvature $C$ is a hermitian 2-form with {\it
homogeneous} gauge transformations
 \bb   C^g :=
\delta(H^g)+(H^g)^2 = \rho(g) C \rho(g^{-1}).\eee
We define the ``preliminary Higgs potential'' $V_0(H)$,
a functional on the  space of Higgses, by
\bb   V_0(H)
:= (C,C) = \t[(\delta H+H^2)P(\delta H+H^2)].\eee
It is a
polynomial of degree 4 in $H$ with real, non-negative
values.  Furthermore it is gauge invariant,
$    V_0(H^g)= V_0(H)$
 because of the homogeneous transformation
property of the  curvature $C$ and because the
orthogonal projector $P$ commutes  with all gauge
transformations
$\rho(g)P =P\rho(g)$.
The
transformation law for $H$, equation (\ref{gauge}),
motivates the following change of  variables
\bb   \Phi := H-i\dd.\label{Phi}\ee
The new variable $\Phi$ transforms
homogeneously
\bb   \Phi^g = \rho(g)\Phi\rho(g^{-1})\eee
 where
the differential is, of course, considered gauge
invariant  $\dd^g = \dd$.

The central result of Connes' scheme is the following:
Let us repeat the procedure outlined above where we
tensorize the (finite dimensional) internal algebra
$\aa$ with the (infinite dimensional) algebra of
functions on spacetime and where we tensorize the
internal Dirac operator with the genuine Dirac
operator. Then the Higgs (in the now infinite
dimensional space of 1-forms) consists of a multiplet of
scalar {\it fields} $H(x)$ and a genuine $G$-gauge
potential, i.e. a differential 1-form valued in the Lie
algebra of the group of unitaries $G$ of the internal
algebra. Furthermore after a suitable regularisation of
the trace, (at this point spacetime has to be supposed
compact and with Euclidean signature), the
preliminary Higgs potential in the infinite
dimensional space can be computed and it is the
complete bosonic action of a Yang-\-Mills-\-Higgs
model with, in general, spontaneously broken gauge
symmetry $G$:
\bb \int\t F^{\mu\nu}F_{\mu\nu}\,\sqrt g\, {\rm d}^4x
+\int\t D^\mu\Phi^*D_\mu\Phi\,\sqrt g\, {\rm d}^4x
+\int V(H)\,\sqrt g\, {\rm d}^4x\eee
where (full) Higgs potential is given by
\bb V(H)=V_0(H)-<\alpha C,\alpha C>
=\t \lb(C-\alpha C)^2\rb.\eee
Here $\alpha$ is the
linear map
\bb   \alpha: \Omega^2\aa\longrightarrow
          \lb\rho(\aa)+\pi(\delta(\ker\pi)^1)\rb^\cc\eee
determined by the two equations
\bb
<R,C-\alpha C>&=&0\qquad{\rm for\ all}\
R\in\rho(\aa)^\cc, \label{a1}\\
 <K,\alpha C>&=&0\qquad {\rm for\ all}\
K\in\pi(\delta(\ker\pi)^1)^\cc.\label{a2}\ee
The scalar product
is the finite dimensional one in
$\pi\lp\hat\Omega^2\aa\rp$,
the $x$ dependence of $C$ can be ignored.
Consequently the (full) Higgs potential is still a
non-negative, invariant polynomial of degree 4.

\section{A $U(2)$ model}

We choose as
internal algebra $\aa=M_2(\cc)$, the algebra of
complex $2\times 2$ matrices. Both left and right
handed fermions come in $N$ generations of doublets,
i.e. the fermions are elements of the Hilbert space
\bb \hh\ :=\ \hh_L\oplus\hh_R \:=\
\cc^2\otimes\cc^N\ \oplus\ \cc^2\otimes\cc^N.\nonumber\ee
This Hilbert space carries the representation
\bb \rho(a)=\pmatrix{
\rho_L(a)&0\cr
0&\rho_R(a)}=\pmatrix{
a\otimes 1_N&0\cr
0&a\otimes 1_N}, \quad a\in\aa\nonumber\ee
The internal Dirac operator is
\bb \dd:=\pmatrix{
0&M\cr
M^*&0},\nonumber\ee
where we choose the fermion mass matrix of block
diagonal form
\bb M=\pmatrix{
m_1&0\cr
0&m_2}=e\otimes m_1+(1-e)\otimes m_2=
e\otimes\mu+1\otimes m_2,\nonumber\ee
 $m_1$ and $m_2$ are complex $N\times N$
matrices which should be thought of as mass matrices
of the quarks of electric charge 2/3  and -1/3 and we
suppose them different, $m_1\neq m_2$. The total mass
matrix $M$ is chosen block diagonal to ensure
conservation of electric charge. We introduced the
shorthands \bb e: =\pmatrix{
1&0\cr
0&0},\nonumber\ee
\bb\mu:=m_1-m_2\neq 0.\nonumber\ee
We have to compute $\Omega^1\aa$ and
$\Omega^2\aa$. In degree 1, we have
\bb \pi(a_0\delta a_1)&=&(-i)\pmatrix{
0&\rho_L(a_0)[M\rho_R(a_1)-\rho_L(a_1)M]\cr
\rho_R(a_0)[M^*\rho_L(a_1)-\rho_R(a_1)M^*]&0}\cr
&=&i\pmatrix{
0&a_0[a_1e-ea_1]\otimes\mu\cr
a_0[a_1e-ea_1]\otimes\mu^*&0}\nonumber\ee
and in degree 2
\bb \pi(a_0\delta a_1\delta a_2)=i^2\pmatrix{
a_0[a_1e-ea_1][a_2e-ea_2]\otimes\mu\mu^*&0\cr
0&a_0[a_1e-ea_1][a_2e-ea_2]\otimes\mu^*\mu}.\nonumber\ee
Let us now show that
\bb \pi(\delta(\ker\pi)^{1}) =\{0\}.\nonumber\ee
In fact, a general element in
$\pi(\delta(\ker\pi)^{1})$ consists of a sum of matrices
of the form
\bb i^2\pmatrix{
[a_0e-ea_0][a_1e-ea_1]\otimes\mu\mu^*&0\cr
0&[a_0e-ea_0][a_1e-ea_1]\otimes\mu^*\mu}
\label{gen}\ee
with the constraint
\bb a_0[a_1e-ea_1]\otimes\mu=
a_0[a_1e-ea_1]\otimes\mu^*=0.\label{con}\ee
Multiplication of the first term in equation (\ref{con})
by $1\ot\mu^*$ on the right and by $e\ot 1$ on the left
shows that the upper left coefficient in (\ref{gen})
vanishes and by symmetry the entire matrix
(\ref{gen}) is zero. In a similar fashion, the images of
all higher kernels are shown to vanish. Therefore, we
have in this example
\bb \Omega\aa=\pi(\hat\Omega\aa)\nonumber\ee
to be contrasted to the vector-like model $m_1=m_2=1$
where $\Omega\aa$ vanishes in all positive degrees.

To compute the Higgs potential, we need an explicit
expression of the differential $\delta$ in degree zero
and one:
\bb\delta\pmatrix{
a\otimes 1&0\cr
0&a\otimes 1}=i\pp{
0&[ae-ea]\otimes\mu\cr
[ae-ea]\otimes\mu^*&0},\nonumber\ee
\bb\delta\pp{
0&ih\otimes\mu\cr
ih\otimes\mu^*&0}=\pp{
[ehe-(1-e)h(1-e)]\otimes\mu\mu^*&0\cr
0&[ehe-(1-e)h(1-e)]\otimes\mu^*\mu}.\nonumber\ee

A Higgs is given by
\bb H=i\pp{ 0&h\otimes\mu\cr
h\otimes\mu^*&0}\in\Omega^1\aa, \quad h=h^*.\nonumber\ee
Its curvature is the hermitian 2-form
\bb   C := \delta H+H^2=:\pp{
c\otimes\mu\mu^*&0\cr
0&c\otimes\mu^*\mu} \in
\Omega^2\aa\nonumber\ee
with
\bb c=ehe-(1-e)h(1-e)-h^2.\nonumber\ee
In our example the gauge group is
\bb    G = \{g\in
\aa,\ gg^\ast =g^*g=1\}=U(2) .\nonumber\ee
Under a gauge transformation $g$, the Higgses
transform inhomogeneously
\bb H^g &:=&\
\rho(g)H\rho(g^{-1})+\rho(g)\delta \rho(g^{-1}) \cr
       &=&\ \rho(g)H\rho(g^{-1})+(-i)\rho(g)[\dd,\rho
(g^{-1})] \cr
       &=:& i\pmatrix {0&h^g\otimes\mu\cr
h^g\otimes\mu^*&0}\nonumber\ee
 with
\bb    h^g = g(h-e)g^{-1} + e\nonumber\ee
while the curvature transforms homogeneously
 \bb   C^g :=
\delta(H^g)+(H^g)^2 = \rho(g) C \rho(g^{-1}):=\pp{
c^g\otimes\mu\mu^*&0\cr
0&c^g\otimes\mu^*\mu}\nonumber\ee
 with
\bb c^g=gcg^{-1}.\nonumber\ee
With the change of variables, equation (\ref{Phi})
 \bb   \Phi := i\pmatrix {0&\phi\otimes
\mu\cr \phi\otimes\mu^*&0} := H-i\pp{
0&e\otimes\mu\cr
e\ot\mu^*&0},\nonumber\ee
we get
 \bb   \phi =h-e, \quad {\rm and}\quad
\phi^g = g\phi g^{-1}.\eee
   In this example the
(translated) Higgses sit in the adjoint representation of
the gauge group. In terms of  $\phi$,  the curvature
becomes
\bb
c=ehe-(1-e)h(1-e)-h^2=-\phi(1+\phi)\nonumber\ee
and the preliminary Higgs potential reads \bb
V_0(H)=2\t\left((\mu\mu^*)^2\right)\t\lb\phi^2
(1+\phi)^2\rb.\nonumber\ee Here, with
$\pi(\delta(\ker\pi)^1) =\{0\},$ $\alpha$ is the linear
map   \bb   \alpha:\Omega^2\aa\longrightarrow
          \rho(\aa)\nonumber\ee
determined by the equation
\bb
<R,C-\alpha C>&=&0\qquad{\rm for\ all}\
R\in\rho(\aa).\nonumber\ee
Therefore
\bb \alpha C=\frac{\t\lp\mu\mu^*\rp}{N}\pp{
c\ot 1&0\cr
0&c\ot 1}\nonumber\ee
and the Higgs potential is
\bb
V(H)=2\lp\t\lp(\mu\mu^*)^2\rp-\frac{(\t\mu\mu^*)
^2}{N}\rp\t\lb\phi^2(1+\phi)^2\rb.\nonumber\ee  It has two
minima, $\phi=0$ and $\phi=-1$. Both are gauge
invariant and the gauge group is not broken
spontaneously. Note that for one generation, $N=1$, the
Higgs potential vanishes identically.

\section{$G=U(2)_L\times U(2)_R$}

Now our internal algebra is
\bb \aa=M_2(\cc)\op M_2(\cc)\ \ni (a,b).\nonumber\ee
The fermions live in $N$ generations of doublets
\bb  \hh\ :=\ \hh_L\oplus\hh_R \:=\
\cc^2\otimes\cc^N\ \oplus\ \cc^2\otimes\cc^N,\nonumber\ee
with representation
\bb \rho(a,b)=\pmatrix{
\rho_L(a)&0\cr
0&\rho_R(b)}=\pmatrix{
a\otimes 1_N&0\cr
0&b\otimes 1_N}, \quad (a,b)\in\aa.\nonumber\ee
Internal Dirac operator and mass matrix are as in
section 1:
 \bb \dd:=\pmatrix{
0&M\cr
M^*&0},\nonumber\ee
\bb M=\pmatrix{
m_1&0\cr
0&m_2}=e\otimes m_1+(1-e)\otimes m_2.\nonumber\ee
We assume that $m_2$ is not a multiple of $m_1$.
In order to compute $\pi(\hat\Omega^p\aa)$, we need
the commutator
\bb  [\dd,\rho(a,b)]=
\pmatrix
{0&M\rho_R(b)-\rho_L(a) M\cr M^*\rho_L(a)-
\rho_R(b)
M^*&0}=\eee
$$\pp{
0&[eb-ae]\ot m_1+[(ae-be)-(a-b)]\ot m_2\cr
[ea-be]\ot m_1^*+[(be-ae)-(b-a)]\ot m_2^*&0}.$$
We find in degree 1
\bb
\pi\lp\delta(a,b)\rp=\eee
$$i\pp{ 0&(ae-eb)\ot m_1+(a(1-e)-(1-e)b)\ot m_2\cr
(be-ea)\ot m_1^*+(b(1-e)-(1-e)a)\ot m_2^*&0},$$
\bb \pi\lp(a_0,b_0)\delta(a_1,b_1)\rp&=&
\rho(a_0,b_0)\pi\lp\delta(a_1,b_1)\rp\cr
&=&
i\pp{
0&h_1\ot m_1+h_2\ot m_2\cr
\tilde h_1\ot m_1^*+\tilde h_2\ot m_2^*&0}
\label{h}\ee
with
\bb h_1&=&a_0(a_1e-eb_1)\cr
h_2&=& a_0[(a_1-b_1)-(a_1e-eb_1)].\nonumber\ee
The $\tilde h_j$ are obtained from the $h_{j}$ by
interchanging $a$'s and $b$'s.
In degree 2, we have
\bb \pi\lp \delta(a_0,b_0)\delta(a_1,b_1)\rp&=&
\pi\lp \delta(a_0,b_0)\rp\pi\lp\delta(a_1,b_1)\rp\cr
&=&\pp{
\sum_{j,k=1}^{2}x_{jk}\ot m_jm_k^*&0\cr
0&\sum_{j,k=1}^{2}\tilde x_{jk}\ot m_j^*m_k}\nonumber\ee
with

\bb x_{11}&=&(a_0e-eb_0)(ea_1-b_1e),\label{i}\\
x_{12}&=&(a_0e-eb_0)\lb(a_1-b_1)-(ea_1-b_1e)\rb,
\\
x_{21}&=&\lb(a_0-b_0)+(a_0e-eb_0)\rb(ea_1-b_1e),\\
x_{22}&=&\lb(a_0e-eb_0)-(a_0-b_0)\rb(ea_1-b_1e)\cr
               &&+\lb(a_0-b_0)-(a_0e-eb_0)\rb(a_1-b_1).
\label{iiii}\ee
The $\tilde x_{jk}$ are obtained from the $x_{jk}$ by
interchanging $a$'s and $b$'s. Trying to rewrite these
formulas in terms of the $h$'s, we get
\bb x_{11}+x_{22}&=&h_1e+h_2(1-e)+e\tilde h_1
+(1-e)\tilde h_2,\cr
x_{12}&=&h_1(1-e)+e\tilde h_2,\cr
x_{21}&=&h_2e+(1-e)\tilde h_1,\nonumber\ee
but
\bb  x_{11}=a_0ea_1-a_0a_1e+(h_1+h_2)e+e\tilde
h_1\nonumber\ee
cannot be expressed in terms of the $h$'s entirely
because of the term $a_0ea_1-a_0a_1e$ and similarly
for $x_{22}$.

To write
\bb MM^*=1\ot\Sigma+\pp{1&0\cr0&-1}\ot\Delta\nonumber\ee
and
\bb
M^*M=1\ot\tilde\Sigma+
\pp{1&0\cr0&-1}\ot\tilde\Delta,\nonumber\ee
we introduce the following shorthands
\bb \Sigma &:=&   {1\over2}(m_1m_1^*+m_2m_2^*),\cr
\Delta & :=&   {1\over2}(m_1m_1^*-m_2m_2^*),\cr
\tilde\Sigma &:=&
{1\over2}(m_1^*m_1+m_2^*m_2),\cr
\tilde\Delta &:=& {1\over2}(m_1^*m_1-m_2^*m_2).\nonumber\ee
We are now ready to compute the junk
$J^2:=\pi(\delta(\ker\pi)^{1})$. The 1-form in
equation (\ref{h}) vanishes if and only if $h_1=h_2=
\tilde h_1=\tilde h_2=0$,
because by assumption $m_2$ is not a multiple of
$m_1$. In this case we have by equations (\ref{i}-%
\ref{iiii})
\bb x_{11}=-x_{22}=a_0ea_1-a_0a_1e\quad
{\rm and}\quad
x_{11}=x_{22}=0.\nonumber\ee
Thus,
 \bb  J^2&=&\la\pp{
\sum_ix_{jk}^i\ot m_jm_k^*&0\cr
0&\sum_i\tilde
x_{jk}^i\ot m_j^*m_k},\ \sum_ih_1^i=\sum_ih_2^i=
\sum_i\tilde h_1^i= \sum_i\tilde h_2^i=0\ra\cr
&\supset&\la\pp{\sum_i a_0^iea_1^i\ot\Delta&0\cr
0&\sum_i b_0^ieb_1^i\ot\tilde\Delta},\
\sum_i a_0^ia_1^i=\sum_i b_0^ib_1^i=0\ra\cr
&=&\la\pp{
x\ot \Delta&0\cr
0&\tilde x\ot\tilde\Delta},\ x,\tilde x\in
M_2(\cc)\ra. \label{sbs}\ee
To prove the last equality in equation (\ref{sbs}),
we note that the subspace is a two-\-sided  ideal in the
rhs. Furthermore the subspace contains non-zero
elements, for instance if:
\bb   a_0 := \pmatrix {0&0\cr 1&-1} \quad {\rm and}
\quad   a_1
:= \pmatrix {1&1\cr 1&1},\nonumber\ee
then $   a_0a_1 = 0$ and  $a_0e a_1 \neq 0$.    The
algebra $M_2(\cc)$ being simple, the subspace
coincides with the whole algebra. Consequently
\bb   J^2=\la\pp{
x\ot \Delta&0\cr
0&\tilde x\ot\tilde\Delta},\ x,\tilde x\in
M_2(\cc)\ra.\nonumber\ee

We need an explicit expression for the orthogonal
projector on the ortho-\-complement of $J^2$ in
$\pi\lp\hat\Omega^2\aa\rp$:
\bb P\pp{
x_{jk}\ot m_jm_k^*&0\cr
0&\tilde x_{jk}\ot m_j^*m_k}\nonumber\ee
$$=\ \pp{
(x_{11}+x_{22})\ot\Sigma'+x_{12}\ot m_{12}'+x_{21}\ot
m_{21}'&0\cr
 0&(\tilde x_{11}+\tilde x_{22})\ot\tilde \Sigma'+\tilde
x_{12}\ot \tilde  m_{12}'+\tilde x_{21}\ot \tilde m_{21}'
}$$
with
\bb m_{12}:=m_1m_2^*,\qquad
m_{21}:=m_2m_1^*.\nonumber\ee
The corresponding expressions with tildes are obtained
from the ones without by tilding all $h$'s or by
interchanging all $m_j$ and $m_j^*$.
The prime denotes projecting out $\Delta$ and
$\tilde\Delta$:
\bb\bullet'&:=&\bullet- \frac{\t(\bullet\Delta)}
{\t(\Delta^2)}\Delta,\cr
\tilde\bullet'&:=&\tilde\bullet- \frac{\t(\tilde\bullet
\tilde\Delta)}
{\t(\tilde\Delta^2)}\tilde\Delta.\nonumber\ee
 Now as operator in
$\Omega^2\aa=P\pi(\hat\Omega^2\aa)$, the derivation
$\delta:\ \Omega^1\aa\rightarrow\Omega^2\aa $ takes
the following form :
\bb
\delta\pi\lp(a_0,b_0)\delta(a_1,b_1)\rp=iP\delta\pp{
0&h_j\ot m_j\cr \tilde h_j\ot m_j^*&0}\nonumber\ee
$$=\ \pp{
x\ot\Sigma'+x_{12}\ot m_{12}'+x_{21}\ot m_{21}'&0\cr
0&\tilde x\ot\tilde \Sigma'+\tilde x_{12}\ot \tilde
m_{12}'+\tilde x_{21}\ot \tilde m_{21}' }$$
where
\bb
x&:=&h_1e+e\tilde h_1+h_2(1-e)+(1-e)\tilde h_2,\cr
x_{12}&:=&h_1(1-e)+e\tilde h_2,\cr
x_{21}&:=&h_2e+(1-e)\tilde h_1.\nonumber\ee

The Higgs is in
$\Omega^1\aa$
\bb H=i\pp{
0&h_1\ot m_1+h_2\ot m_2\cr
 h_1^*\ot m_1^*+ h_2^*\ot m_2^*&0}.\nonumber\ee
Its curvature is the 2-form in $\Omega^2\aa$
\bb
 C:= \delta H+H^2\eee
$$=\ \pp{
c\ot\Sigma'+c_{12}\ot m_{12}'+c_{21}\ot m_{21}'&0\cr
0&\tilde c\ot\tilde \Sigma'+\tilde c_{12}\ot \tilde
m_{12}'+\tilde c_{21}\ot \tilde m_{21}' }$$
with
\bb c&:=&h_1e+eh_1^*+h_2(1-e)+(1-e)h_2^*,\cr
c_{12}&:=&h_1(1-e)+eh_2^*,\cr
c_{21}&:=&h_2e+(1-e)h_1^*.\nonumber\ee
Now the expressions with tildes are obtained
from the ones without upon replacing all $h_j$ by
$h_j^*$.

Under a gauge transformation
\bb g=(g_L,g_R)\ \in\
 G= \la g\in \aa,\ gg^\ast
=g^*g=1\ra=U(2)_L\times U(2)_R\nonumber\ee
the Higgs transforms as
\bb H^g=i\pp{
0&h_1^g\ot m_1+h_2^g\ot m_2\cr
 h_1^{g*}\ot m_1^*+ h_2^{g*}\ot m_2^*&0}\nonumber\ee
with
\bb h_1^g&=&g_Lh_1g_R^{-1}-g_Leg_R^{-1}+e,\cr
h_2^g&=&g_Lh_2g_R^{-1}-g_L(1-e)g_R^{-1}+(1-e).\nonumber\ee
Again, we pass to the homogeneous Higgs variables
\bb\phi_1&:=&h_1-e,\cr
\phi_2&:=&h_2-(1-e),\cr
\phi^g_j&=&g_L\phi_jg_R^{-1},\quad j=1,2\nonumber\ee
and
\bb\Phi=i\pp{
0&\phi_1\ot m_1+\phi_2\ot m_2\cr
 \phi_1^*\ot m_1^*+ \phi_2^*\ot m_2^*&0}=H-i\dd.\nonumber\ee
In these variables, the curvature becomes
\bb c&=&1-\phi_1\phi_1^*-\phi_2\phi_2^*,\cr
c_{12}&=&-\phi_1\phi_2^*,\cr
c_{21}&=&-\phi_2\phi_1^*.\nonumber\ee
There is one and only one gauge invariant point in
the space of Higgses namely $\Phi=0$. The curvature of
this point is different from zero because $c=1$. The
preliminary Higgs potential is

\begin{samepage}
\bb V_0(H)&=&(C,C)\cr
&=&2[
\t\lp c^2\rp\t\lp\Sigma'^2\rp+
\t\lp c_{12}^2\rp\t\lp m_{12}'^2\rp+
\t\lp c_{21}^2\rp\t\lp m_{21}'^2\rp\cr
&&\quad 2\t\lp cc_{12}\rp\t\lp\Sigma'm'_{12}\rp
+2\t\lp cc_{21}\rp\t\lp\Sigma'm'_{21}\rp
+2\t\lp c_{12}c_{21}\rp\t\lp m'_{12}m'_{21}\rp]\eee
\end{samepage}
and breaks the gauge symmetry spontaneously. The
Higgs potential
\bb V(H)= V_0(H)\,-<\alpha C,\alpha C>\,=\t\lb\lp
C-\alpha C \rp^2\rb\nonumber\ee
is computed with the linear map
\bb   \alpha: \Omega^2\aa\longrightarrow
          \rho(\aa)+\pi(\delta(\ker\pi)^1)\nonumber\ee
determined by the two equations (\ref{a1},\ref{a2}).
A straightforward calculation yields
\bb \alpha C=\pp{
\lb c\t\Sigma'+c_{12}\t m'_{12}+c_{21}\t m'_{21}\rb
\ot Y&0\cr
0&\lb \tilde c\t\Sigma'+\tilde c_{12}\t m'_{12}+
\tilde c_{21}\t
m'_{21}\rb \ot\tilde Y}\nonumber\ee
with
\bb Y:=\frac{1}{N-\lp\t\Delta\rp^2/\t\lp\Delta^2\rp}
\lp 1-\frac{\t\Delta}{\t\lp\Delta^2\rp}\Delta\rp.\nonumber\ee
Therefore also the Higgs potential breaks the gauge
symmetry spontaneously (unless there is a numerical
accident in the fermionic mass matrix). The
vacuum expectation value of  $\Phi$ is any point in
the orbit of $\Phi-H=-i\dd$, for instance
\bb\phi_1=e,\qquad\phi_2=1-e,\eee
and the left handed gauge bosons acquire the same
masses as the right handed ones.
Indeed, parity remains unbroken because the
Higgs representation consists of two complex
$(2_L,2_R)$, $\phi_1$ and $\phi_2$ [5].

\section{Conclusion}

The main motivation of this work was to find a
Connes-\-Lott
model, that enjoys spontaneous
breaking of gauge symmetry and parity
simultaneously. This hope was spoiled. In a
general left-\-right symmetric model, e.g. $\aa=
M_3(\cc)\op M_3(\cc)$, we are unable to compute
explicitly the junk and the differential
$\delta$ from $\Omega^1\aa$ to $\Omega^2\aa$ and we
are therefore unable to decide whether the gauge
symmetry is broken or not. Nevertheless, it is pretty
clear that any vacuum expectation, that might come
out, will be an element of $\rho_L\ot\rho_R$ and
parity preserving.

Chamseddine \& Fr\"ohlich [6] have considered
the left-\-right symmetric, grand unified $SO(10)$
model in the Connes-\-Lott setting. Without computing
the junk $J^2$, they also conclude that
parity breaking does not occur.

There are two alternative algorithms applying
non-\-commutative geometry to particle physics. One
is due Dubois-Violette, Madore \& Kerner [7]. In their
scheme the differential algebra $\Omega\aa$ is defined
in terms of derivations and does not depend on
fermion representations. The other algorithm, due to
Coquereaux [8], takes the
differential algebra as starting point and is thereby
more flexible. Both algorithms also yield
spontaneous break down of gauge symmetry and it
would be interesting to know if they can accommodate
spontaneous parity violation.

It is a pleasure to acknowledge Pierre Bin\'etruy's
advice.

 \vfil\eject

\centerline{\bf References}

\begin{description}
\item{\ [1]} A. Connes, {\it Non-Commutative
Geometry},
Publ. Math. IHES 62 (1985),\hfil\break
A. Connes \& J. Lott, Nucl. Phys. Proc. Suppl.
B18 (1989) 29, \hfil\break
 A. Connes, {\it Non-Commutative Geometry}, Academic
Press (1993)\\
 A. Connes \& J. Lott, {\it The metric
aspect of non-commutative geometry}, in the
proceedings of the 1991 Carg\`ese Summer Conference,
eds.: J. Fr\"ohlich et al., Plenum Press (1992)
\item{\ [2]} D. Kastler, A detailed account of Alain
 Connes' version of the standard model in
non-commutative differential geometry I, II, and III.
to appear in Rev.Math.Phys. \hfil\break
D. Kastler \& M. Mebkhout, {\it Lectures on
Non-Commutative Differential Geometry}, World
Scientific, to be published \hfil\break
 D. Kastler \& T. Sch\"ucker,
Theor. Math. Phys. 92 (1992) 522
\item{\ [3]} J.C. V\'arilly \&  J. M. Gracia-Bond\'\i a, {\it
Connes' noncommutative differential geometry and
the standard model}, J. Geom. Phys., in press
\item{\ [4]} T. Sch\"ucker \& J.-M. Zylinski, {\it
Connes' model building kit}, CPT-93/P.2960
\item{\ [5]} P. Bin\'etruy, M. K. Gaillard
\& Z. Kunszt, Nucl. Phys. B144 (1978) 141
\item{\ [6]} A. Chamseddine \& J.
Fr\"ohlich, {\it $SO(10)$ Unification in
non-commutative geometry},  ZU-TH-10/1993,
ETH/TH/93-12
\item{\ [6]} M. Dubois-Violette, C. R. Acad. Sc.
Paris 307 I (1988) 403 \hfil\break
M. Dubois-Violette, R. Kerner \& J.
Madore, Phys. Lett. 217B (1989), Class. Quant. Grav. 6
(1989) 1709, J. Math. Phys 31 (1990) 316, J. Math. Phys.
31 (1990) 323, Class. Quant. Grav. 8 (1991) 1077,
\hfil\break H. Grosse \& J. Madore, Phys. Lett. 283B
(1992) 218, \hfil\break
J. Madore, Mod. Phys. Lett. A4 (1989) 2617,
J. Math. Phys. 32 (1991) 332,
Int. J. Mod. Phys. A6 (1991) 1287,
Phys. Lett. 305B (1993) 84,
{\it On a non-commutative
extension of Electro\-dynamics}, LPTHE Orsay 92/21
\hfil\break
B. S. Balakrishna, F. G\"ursey \&
K. C. Wali, Phys. Lett. 254B (1991) 430, Phys. Rev. D 44
(1991) 3313
\item{[8]} R. Coquereaux, G. Esposito-Far\`ese \&
G. Vaillant, Nucl. Phys.B353 (1991) 689
\hfil\break
R. Coquereaux, G. Esposito-Far\`ese \&
F. Scheck, Int. J. Mod. Phys. A7 (1992) 6555,
\hfil\break
R. H\"au\ss ling, N.A. Papadopoulos \& F. Scheck,
Phys. Lett. 260B (1991) 125,
Phys. Lett. 303B (1993) 265 \hfil\break
R. Coquereaux, R. H\"au\ss ling, N.A. Papadopoulos
\& F. Scheck, Int. J. Mod. Phys. A7 (1992) 2809
\hfil\break
F. Scheck, Phys. Lett. 284B (1992) 303,
\hfil\break
R. Coquereaux, R. H\"au\ss ling \& F. Scheck, {\it Algebraic
Connections on Parallel Universes}, preprint
Universities of Macquarie and Mainz 1993
\hfil\break
N.A. Papadopoulos, J. Plass \& F. Scheck, {\it Models of
elctroweak interactions in non-\-commutative
geometry: a comparison},
 MZ-TH/93-26

\end{description}
\vfil\eject
\end{document}